\documentclass[a4paper]{jpconf}
\usepackage{graphicx}
\newcommand{\avg}[1]{\langle #1\rangle}
\newcommand{\histoAvg}[1]{ \overline{\{#1 \} } }

\begin{document}
\title{Femtoscopy and energy-momentum conservation effects in proton-proton collisions at 900 GeV in ALICE}

\author{Nicolas Bock}

\address{Department of Physics, The Ohio State University, 191 W Woodruff Ave, Columbus Ohio 43212, USA}

\ead{bock@mps.ohio-state.edu}

\begin{abstract}
Two particle correlations are used to extract information about the characteristic size of the system
for proton-proton collisions at $\sqrt{s}$=900 GeV measured by the ALICE (A Large Ion Collider Experiment) detector at CERN. The correlation functions obtained show the expected Bose-Einstein effect for identical particles, but there are also long range correlations present that shift the baseline from the expected flat behavior. A possible source
of these correlations is the conservation of energy and momentum, especially for small systems, where the energy
available for particle production is limited. A new technique, first introduced by the STAR collaboration,  of quantifying
these long range correlations using energy-momentum conservation considerations is presented here. It is shown that 
the baseline of the two particle correlation function can be described using this technique. 

\end{abstract}

\section{Introduction}
\begin{figure}[h]
\begin{minipage}{14pc}
\includegraphics[width=18pc]{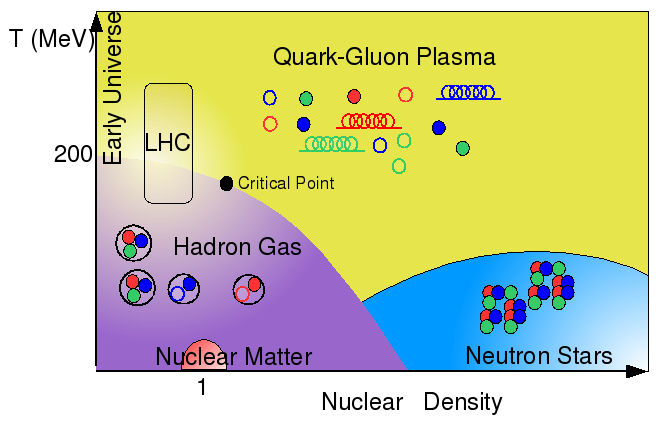}\hspace{2pc}%
\caption{\label{QCD}The QCD phase diagram shows the different phases of nuclear matter.}
\end{minipage}\hspace{7pc}%
\begin{minipage}{14pc}
\includegraphics[width=18pc, height = 10pc]{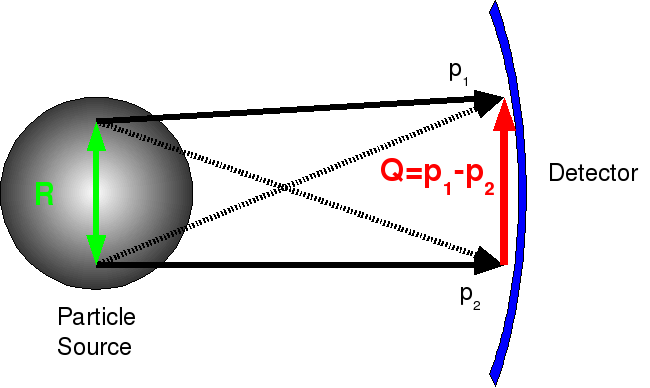}\hspace{2pc}%
\caption{\label{HBTFig} Two identical particles are emitted and detected simultaneously. There are two possible ways in which the particles could have been emitted.}
\end{minipage} 
\end{figure}

Since November 2009 the Large Hadron Collider (LHC) has produced 14M proton-proton collisions at $\sqrt{s}$=900 GeV in ALICE (A Large Ion Collider Experiment) \cite{Alice}.The  ALICE detector is designed to investigate matter at the highest energy densities, where it is supposed to be deconfined. This state of matter is composed of free quarks and gluons, therefore known as the Quark Gluon Plasma (QGP), and it is believed that right after the Big Bang matter was in this state. The QCD phase diagram (Fig.~\ref{QCD}) shows
the different phases of nuclear matter, and it is of great interest to study the properties of the QGP like the temperature of the phase transition, 
the critical point, the energy density, time evolution, lifetime of the system and flow-like effects. 
The present study is concerned with 
the measurement of the size of the particle emission region in proton proton collisions at $\sqrt{s}$=900 GeV using femtoscopy, and how it is affected by energy momentum conservation effects.

\section{Femtoscopy in ALICE}
\label{femto}
Femtoscopy \cite{Femto}, or HBT \cite{HBT}, is a technique that measures the shape and size of the particle emission region by measuring
pairs of identical particles simultaneously, see Fig.~\ref{HBTFig}. Pions are most commonly used in this analysis, and because they are indistinguishable (bosons) the two-particle wave function has to be symmetrized as follows:
\begin{equation}
	\Psi(p_1,r_1,p_2,r_2)\sim e^{i (p_1\cdot r_1+ p_2 \cdot r_2)} +  e^{ i (p_1\cdot r_2+ p_2 \cdot r_1)}
\end{equation}
The probability of detecting two such particles is :
\begin{equation}
	|\Psi^2|\sim  1+\cos((p_1-p_2)\cdot(r_1-r_2)) = 1+\cos(Q\cdot \Delta r),
\end{equation}
where $Q=p_1-p_2$ is the  momentum difference and $\Delta r = r_1-r_2$ is the relative emission point of the particles (Fig.~\ref{HBTFig}). Pairs with small relative momentum  $Q$ will be observed with higher rates because Bose-Einstein particles are likely to be found in the same state.  

\begin{center}
\begin{figure}[h]
\begin{minipage}{14pc}
\includegraphics[width=18pc, height= 10pc]{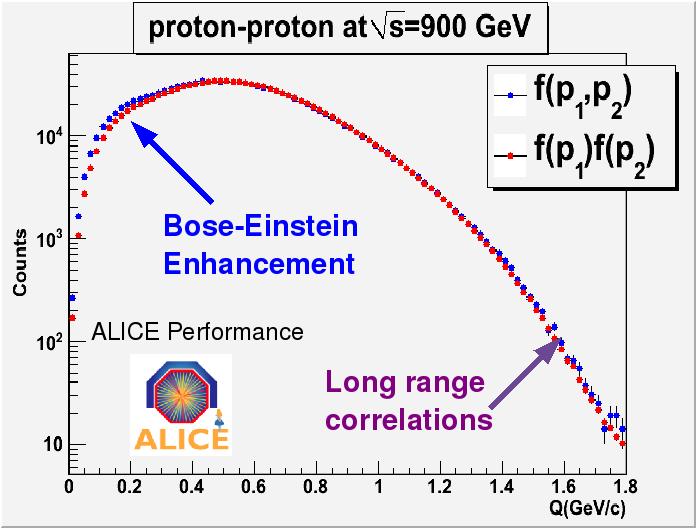}
\caption{\label{NumDen}Two particle distributions for real and mixed events.}
\end{minipage}\hspace{7pc}%
\begin{minipage}{14pc}
\includegraphics[width=18pc, height= 10pc]{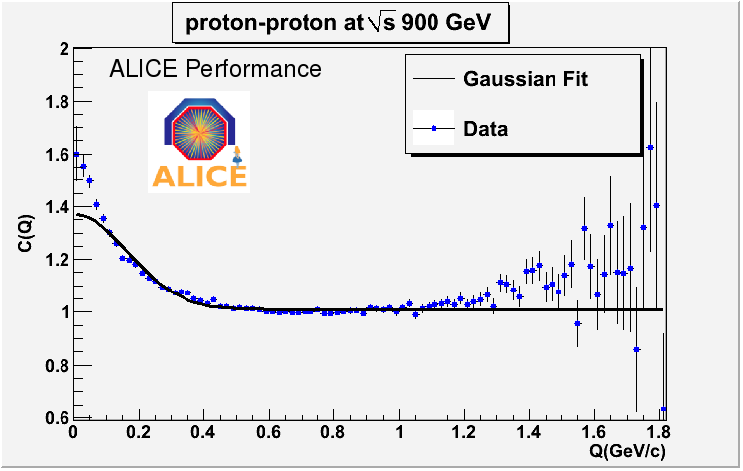}
\caption{\label{CFGauss}Correlation function with a Gaussian fit.}
\end{minipage} 
\end{figure}
\end{center}

For a source of particle emission $\rho(r)$ a correlation function can be obtained by integrating over
all possible pairs of pions resulting in the Fourier transform of the source function:
\begin{equation}
          C_{th}(Q)= \int P(Q)\rho(r_1)\rho(r_2)dr_1 dr_2 = 1+ \hat{\rho}(Q)   
\end{equation}
In particular for a Gaussian source $\rho(r) \sim \exp^{- r^2 / 2R^2}$ one obtains:
\begin{equation} 
\label{gaussian}
          C_{th}(Q)=1+\lambda e^{-R^2Q^2},
\end{equation}
where $\lambda$ is the correlation strength. This equation is used to fit the experimental correlation 
function, defined as the ratio of the two particle and single particle distributions:
\begin {equation}	
          C_{exp}(Q)=\frac{f(p_1,p_2)}{f(p_1)f(p_2)}
\end{equation}
The two particle distribution is obtained by correlating pions from the same event, and the single particle distribution is obtained from mixed events. The data shown in Fig.~\ref{NumDen} is from 250~k  proton-proton events at $\sqrt{s}$ = 900 GeV recorded by ALICE in December 2009. The higher counts at low $Q$ are the Bose-Einstein enhancement, but there are also long range correlations present, which are caused by non-femtoscopic effects such as jets and energy momentum conservation induced correlations (EMCICs for short).
In Fig.(\ref{CFGauss}) $C_{exp}(Q)$ is plotted together with a Gaussian fit. 
It is evident that the simple Gaussian fit neither describes the low $Q$ enhancement, nor the long range correlations. This implies that the particle emission source for proton-proton collisions is non Gaussian and other functional forms such as exponential or Lorentzian can be used. In this paper we keep using a Gaussian as we are interested in describing the baseline. 

\section{Energy-momentum conservation induced correlations}
\label{EMCICs}
\begin{figure}[h]
\begin{minipage}{14pc}
\includegraphics[width=18pc, height= 10pc]{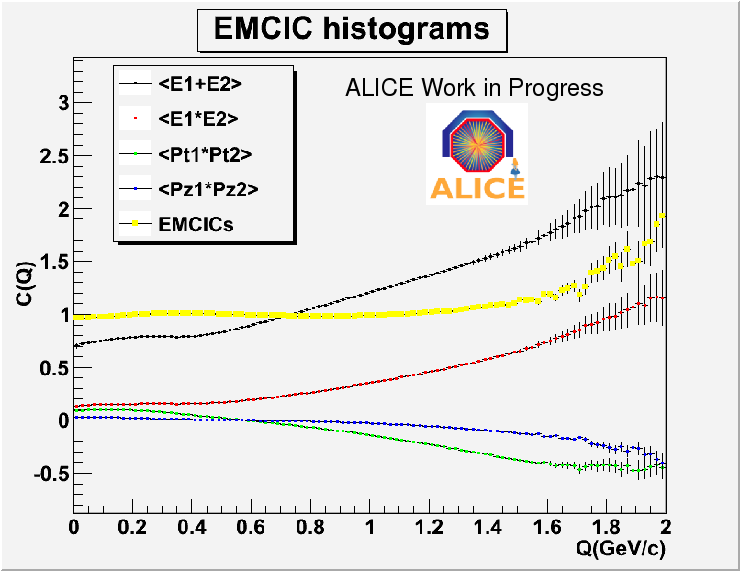}
\caption{\label{EmcicComp}The different EMCIC histograms $\histoAvg{X}$ are shown together with $C_{EMCIC}(Q)$.}
\end{minipage}\hspace{7pc}%
\begin{minipage}{14pc}
\includegraphics[width=18pc, height= 10pc]{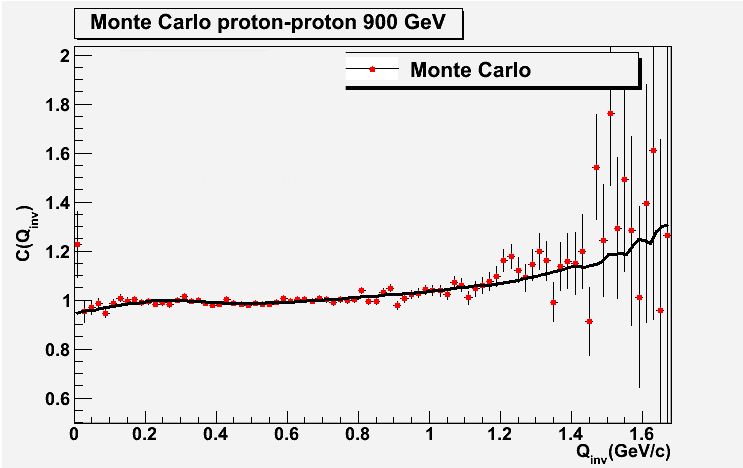}
\caption{\label{MCBase} Eq.~(\ref{Cemcic}) is used to fit Monte Carlo Data.}
\end{minipage} 
\end{figure}

In order to determine the source size better it would be very useful to find a functional form for the sloped baseline observed in Fig.(\ref{CFGauss}). A linear or quadratic slope can be used, but these are ad-hoc parameterizations, and it would be better if the baseline could be determined from first principles.  
It has been shown \cite{Zibi} that particle distributions are affected by EMCICs and if there are no other correlations present the two particle correlation function becomes:
\begin{equation}
\label{emcic1}
C_{EMCIC}(p_1,p_2) = 1 - \frac{1}{N}\Bigg (2 \frac{ \vec{p}_{1,T} \cdot \vec{p}_{2,T}  }{\avg{p_T^2}}
              +\frac{p_{1,z} \cdot p_{2,z} } {\avg{p_z^2}} 
        	+\frac{(E_1-\avg{E})(E_2-\avg{E})   }{\avg{E^2}-\avg{E}^2}\Bigg)  
\end{equation}
where $N$ is the event multiplicity. Equation~\ref{emcic1} characterizes the EMCICs and it would be possible to remove their effect from the experimental correlation function. However, in experiment $N,\avg{p_T^2},\avg{p_z^2},\avg{E}$ and $\avg{E^2}$ cannot be measured, simply because not every single particle is detected. These unknown quantities can be parametrized instead, and the resulting equation can be used as a fit to the data:

\begin{equation}
	\label{Cemcic}
  C_{EMCIC}(p_1,p_2)= \Bigg(1 - M_1\cdot\histoAvg{\vec{p}_{1,T}\cdot \vec{p}_{2,T}} - M_2 \cdot 
             \histoAvg{p_{1,z}\cdot p_{2,z}}-M_3\cdot\histoAvg{E_1\cdot E_2} + 
	M_4\cdot\histoAvg{E_1+E_2}-\frac{M_4^2}{M_3} \Bigg)
\end{equation}
where the $M$ parameters are defined as:
$M_1=\frac{2}{N\avg{p_T^2}}$, $M_2=\frac{1}{N\avg{p_z^2}}$, $M_3=\frac{1}{N(\avg{E^2}-\avg{E}^2)}$, $M_4=\frac{\avg{E}}{N(\avg{E^2}-\avg{E}^2)}$ and the notation $\histoAvg{X}$ represents histograms of two-particle quantities that can be measured in experiment (see Fig~\ref{EmcicComp}). 

Monte Carlo data, which does not have femtoscopic correlations but does guaranty energy and momentum conservation, has been used to test Eq.~(\ref{Cemcic}) as a fitting function for the baseline. Figure \ref{MCBase} shows a good agreement between data and the fitted function. 

\begin{figure}[h]
\includegraphics[width=18pc, height = 10pc]{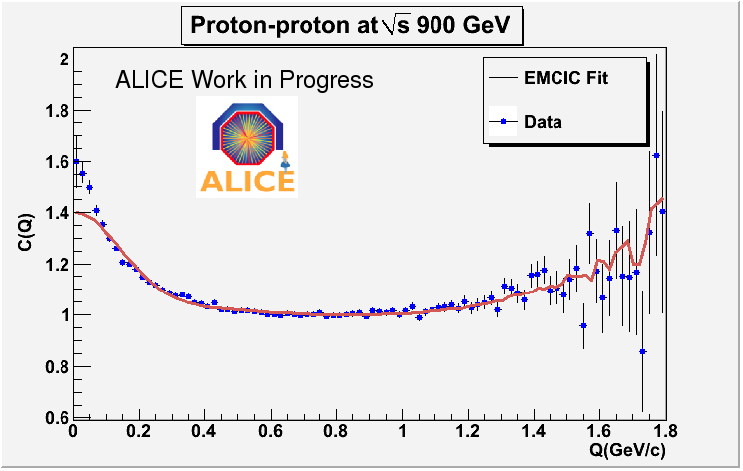}\hspace{2pc}%
\begin{minipage}[b]{14pc}\caption{\label{emcicFit} Fit to the data using Eq.~(\ref{CFtotal}).}
\end{minipage}
\end{figure}

When all correlations are present the total correlation function can be written as:
\begin{equation}
	\label{CFtotal}
   C(Q) = \Phi_{femto}(Q)\times C_{EMCIC}(Q)
\end{equation}
where  $\Phi_{femto} = 1 + \lambda e^{-R^2Q^2}$ is the pure femtoscopic effect. Equation (\ref{CFtotal}) is then 
used as the fitting equation to the real data. The results are shown in Fig.~\ref{emcicFit} and it can be seen
that the baseline is described quite well by this method. The oscillations in the fitting function at large $Q$ arise 
from the limited statistics.

\section{Summary}
Two particle correlations for proton-proton collisions at 900 GeV exhibit long range correlations that 
make the determination of the baseline difficult. These correlations are primarily from energy momentum conservation  
and it has been shown here that they can be quantified and used to determine the baseline correctly. 
The EMCICs introduce more parameters into the fitting function, but the parameterization is obtained from first principles and these preliminary results look promising. 
Future work will be to separate the correlation function and EMCICs in pair average momentum bins $\Big(k_T= \frac{p_1+p_2}{2}\Big)$ and fit all bins simultaneously with one set of parameters $M_1$ through $M_4$, thereby constraining more the parameter space. Three dimensional correlation functions in $Q_{out}, Q_{side}, Q_{long}$ and its decomposition
in spherical harmonics will also be studied.

\ack The author would like to thank Thomas Humanic and Mike Lisa at The Ohio State University for all the 
insight in this field and their continued support and motivation, and the femtoscopy group at the ALICE experiment. 
This work was supported by the U.S. National Science Foundation under grant PHY - 0653432.

\end{document}